\documentstyle[amssymb,pra,aps,epsfig,12pt]{revtex}
\newcommand{\be}{\begin{equation}}
\newcommand{\ee}{\end{equation}}
\newcommand{\br}{\begin{eqnarray}}
\newcommand{\er}{\end{eqnarray}}

\newcommand{\bd}{\begin{displaymath}}
\newcommand{\ed}{\end{displaymath}}

\newcommand{\bfig}{\begin{figure}}
\newcommand{\efig}{\end{figure}}

\def\3cdot{\cdot \cdot \cdot}

\def\om0{\omega _0}
\def\Om0{\Omega _0}

\def\text#1{{\rm{#1}}}

\def\->{\rightarrow}
\def\=>{\Rightarrow}
\def\-->{\longrightarrow}
\def\==>{\Longrightarrow}

\def\pr{^\prime}

\def\pr2{^{\prime\prime}}

\def\bfig{\begin{figure}}
\def\efig{\end{figure}}

\begin{document}
\title{{\Large Decoherence in trapped ions due to polarization of the
residual background gas }}
\author{R. M. Serra$^{1}$\thanks{%
Electronic address: serra@df.ufscar.br}, N. G. de Almeida$^{1}$, W. B. da
Costa$^{2}$, and M. H. Y. Moussa$^{1}$\thanks{%
Electronic address: miled@df.ufscar.br}.}
\address{$^{1}$Departamento de F\'{\i}sica, Universidade Federal de S\~{a}o\\
Carlos, P. O. Box 676, S\~{a}o Carlos, 13565-905, S\~{a}o Paulo, Brazil \\
$^{2}$Universidade Estadual de Mato Grosso do Sul, P. O. Box 351, Dourados,\\
79804-970, Mato Grosso do Sul, Brazil}
\maketitle

\begin{abstract}
We investigate the mechanism of damping and heating of trapped ions
associated with the polarization of the residual background gas induced by
the oscillating ions themselves. Reasoning by analogy with the physics of
surface electrons in liquid helium, we demonstrate that the decay of Rabi
oscillations observed in experiments on $^{9}$Be$^{+}$ can be attributed to
the polarization phenomena investigated here. The measured sensitivity of
the damping of Rabi oscillations with respect to the vibrational quantum
number of a trapped ion is also predicted in our polarization model.

{\bf Journal-Ref:} Physical Review A, {\bf 64}, 033419 (September, 2001)

\end{abstract}

%
%%%%%%%%%%%%%%%%%%%%%%%%%%%%%%%%%%%%%%%%%%%%%%%%%%%%%%%%%%%%%%%%%%%%

\section{Introduction}

The experimental techniques developed over the last decade for manipulating
trapped ions, Rydberg atoms in cavity QED, and traveling fields have enabled
a fruitful dialog to take place between theoretical and experimental
physics, resulting in a mastery of fundamental quantum phenomena at a level
that seems to herald a new phase in the technology of communication \cite%
{comm} and computation \cite{cz,comp}. The improvement of techniques for
generating and detecting nonclassical electronic and vibrational states of
trapped ions and both trapped and traveling nonclassical states of the
radiation field have been of growing interest in a variety of experimental
applications ranging from quantum measurement concepts \cite{gatoH,gatoW} to
teleportation \cite{tele} and quantum logic operations \cite{wine1}. In its
turn, theoretical physics has exploited the possibility of engineering
nonclassical states in some exotic applications such as detection of
gravitational waves \cite{grav}, or for measuring particular properties of
the radiation field, such as its phase \cite{pegg} or its ${\cal Q}$
function \cite{basilio}. Laser manipulation of a string of ions in a linear
trap has been proposed as a way of implementing quantum gates with cold
trapped ions \cite{cz}. Furthermore, interfering laser beams have been used
to induce arrays of microscopic potentials, the optical lattices \cite%
{jaksch}, by the ac Stark effect, opening the way to simulated spin-spin
interactions between trapped atoms, similar to those characteristic of
ferromagnetism and antiferromagnetism in condensed matter physics \cite{SM}.
Quantum computation with ions in thermal motion has also been suggested \cite%
{poyatosCZ,sorensenM}.

The practical realization of such interesting proposals in quantum
communication and computation has come up against the decoherence of quantum
states, owing to the inevitable action of the surrounding environment \cite%
{ANM} and the intrinsic fluctuations in the interaction parameters required
for logic operations \cite{serra}. In fact, the need for huge superpositions
of qubit states for such operations imposes the requirements that the
quantum systems be totally isolated from the environment and that the
interaction parameters involved be tightly controlled. Hence, investigation
of the noise sources in such promising quantum systems turns out to be a
crucial step toward the implementation of a quantum logic processor.

Unlike processes in cavity QED, where decoherence is caused by the
well-known cavity damping mechanisms, spontaneous atomic emission, and
inefficiency of the ionization detectors, in the domain of trapped ions a
large range of error sources have been identified. The effects of
spontaneous emission have been studied \cite{hughes,plenio,djames}, as well
as dephasing due to the ions' zero-point motion \cite{garg}. Alternative
sources of decoherence have been introduced phenomenologically \cite{knight}
and stochastic models have been proposed to handle intensity and phase
fluctuations in the exciting laser pulses, in addition to fluctuations in
the ion trap potential \cite{milburn}. Recently, instead of a stochastic
mechanism, Di Fidio and Vogel \cite{vogeln} have put forward a model in
which the observed damping in Rabi oscillations \cite{gatoW2} is caused by
quantum jumps to an auxiliary electronic level. Other significant sources of
error are the instabilities of the trap drive frequency and voltage
amplitude \cite{gatoW,roos}.

Collisions of the trapped ion with residual background gas (usually H$_{2}$
in the NIST $^{9}$Be$^{+}$ experiments \cite{nist}) can also be an important
error source{\bf , }although experiments are typically carried out in a
high-vacuum environment \cite{nist} at a pressure around $10^{-8}$ Pa. In
connection with this, the current article discusses the polarization effects
on the residual background gas induced by the oscillating trapped ion. We
demonstrate that, apart from other known processes, the damping and heating
mechanism of trapped ions is produced by local polarization of the residual
thermal background gas (BG). Since the density of the BG is rather low,
around $10^{6}$ cm$^{-3}$ \cite{nist}, we do not expect to find a
quasiparticle consisting of the trapped ion together with its surrounding
polarization cloud, the so called polaron. The density of the BG does not
permit a polaron binding energy and the trapped ion is scattered by the BG.
For practical purposes, in our model we assume that the BG is continuous and
that the ion is scattered by its surface oscillations, by analogy with the
interaction of surface electrons in liquid helium (where the oscillations
are called ripplons in their quantized form).

In contrast to loss mechanisms revealed by damped Rabi oscillations, it is
well known that background gas can heat trapped ions by transferring energy
during an elastic collision. A heating rate can easily be estimated from the
total collision cross section \cite{nist}. Although elastic collisons are
expected to be the main source of decoherence in ionic traps, inelastic
collisons also take place, changing the internal state or even the species
of the trapped ion. In experiments on $^{9}$Be$^{+}$, inelastic processes
may convert the ion to BeH$^{+}$ (upon collision with a H$_{2}$ molecule)
when resonant light is applied to the $^{2}$S$_{1/2}\rightarrow $ $^{2}$P$%
_{1/2,3/2}$ transitions \cite{gatoW}. Both types of inelastic collision,
chemical reactions and charge exchange, besides depending critically on the
constituents of the BG, occur only when the interparticle spacing between
the trapped ion and the neutral background atoms approaches atomic
dimensions. In our model, a mean interparticle spacing between the ion and
the BG atoms is introduced. An upper limit on the rate of inelastic
collisions follows from the Langevin rate, which accounts for the background
neutrals that penetrate the angular momentum barrier and undergo a
spiraling-type collision into the ion \cite{nist,hasted}.

The attractive interaction potential resulting from polarization of the
neutral background by the electric field of the trapped ion is given by $%
{\rm U}(r)=-\chi q^{2}/\left( 8\pi \epsilon _{0}r^{4}\right) {\rm ,}$ where $%
\chi $ is the polarizability and $q$ the ionic charge. Spiraling collisions
result when the impact parameter is less than a critical value ${\frak p}%
=\left( \chi q^{2}/\pi \epsilon _{0}{\frak mv}^{2}\right) ^{1/4}$, where $%
{\frak m}$ and ${\frak v}$ are the reduced mass and relative velocity of the
pair. From the critical impact parameter follows the Langevin rate constant $%
{\cal L}=\pi {\frak p}^{2}{\frak v}$, which gives to the reaction rate $%
{\cal R}=\rho {\cal L}=\rho q\left( \pi \chi /\epsilon _{0}{\frak m}\right)
^{1/2}$, where $\rho $ is the density of the BG. From the parameters used in
the NIST $^{9}$Be$^{+}$ experiments, we obtain a small estimated probability
of inelastic collisions with the BG constituents.

In addition to the chemical reactions and charge exchange, the BG can heat
or cool the trapped ion through energy transfer during an elastic collision.
An estimate of the elastic collision rate can be inferred from the total
collision cross section $\sigma $ in a $\Lambda /r^{4}$ potential. Assuming $%
\Lambda ={\rm U}(r)$, it follows, from a conservative estimate, that elastic
collisions will also be rare \cite{nist}. However, the effects of
collisional heating (cooling), as well as the model we present here, can be
tested, as suggested in Ref. \cite{nist}, by raising (lowering) the BG
pressure. On the other hand, it is well known that when ions are first
loaded into a trap, elastic collisions with the BG are beneficial, allowing
laser cooling to proceed faster, the BG providing a viscous medium and
bringing the temperature of the trapped ions into thermal equilibrium with
the surrounding atoms. We conclude from the work reported here that the
damping medium provided by the background has also to be taken into account
when the trapped ions are cooled to their motional ground state. This claim
is supported by the{\bf \ }excellent fit between our model and the
experimental data for measurement of the fluorescence probability of the
electronic ground state, reported in \cite{gatoW2}.

Finally, we mention that, for a linear Paul rf trap, the process of
decoherence may be dominated by that of the motional state, instead of those
due to internal levels, caused by a nonideal applied field \cite{nist}. In
fact, the internal levels of the trapped ion are metastable, typically two
ground state hyperfine sublevels \cite{dav}. While Refs. \cite{milburn} take
into account technical fluctuations in the applied fields used to manipulate
the trapped ions, Ref. \cite{vogeln} attributes the error source to the
coupling of the internal states to the environment. In this paper we present
a model that accounts for the dominant decoherence of the ion motion in
terms of the induced polarization of the BG.

The paper is organized as follows. In Sec. II we present a detailed
discussion of the fundamental ion-laser and ion-BG interactions. In Sec. III
we examine the effects of the ion-BG coupling on the behavior of a prepared
motional-electronic state of the trapped ion, when considering the Carrier
and the anti-Jaynes-Cummings ion-laser interactions. For the latter
ion-laser interactions the Rabi oscillations of a trapped ion initially
cooled to its motional ground state are computed under the ion-BG coupling,
and our results are compared with the available experimental data for a
trapped $^{9}$Be$^{+}$ ion \cite{gatoW2}. Sec. IV is dedicated to comments
and conclusions and, finally, in the Appendix A we show how to obtain a Fr%
\"{o}hlich-type ion-BG interaction, following the model used for surface
electrons in liquid helium.

\section{Ion-laser and ion-BG interactions}

We consider a single trapped ion of mass $m$ in a one-dimensional harmonic
trap of frequency $\nu $. The ion has forbidden transitions between two{\bf %
\ }internal electronic states (excited $|\uparrow \rangle $ and ground $%
|\downarrow \rangle $ states, assumed to be hyperfine sublevels of the
ground state), separated by frequency $\omega _{0}$ and indirectly coupled
by the interaction with two laser beams, with frequencies $\omega _{1}$ and $%
\omega _{2}$, in a stimulated Raman-type configuration. As indicated in Fig.
1, the laser beams are detuned by $\Delta $ from a third more excited level $%
|r\rangle $ which, in the stimulated Raman-type configuration, is
adiabatically eliminated when $\Delta $ is much larger than three
quantities: the linewidth of level $|r\rangle $, the coupling associated
with the $|\uparrow \rangle $ $\leftrightarrow $ $|r\rangle $ and $%
|\downarrow \rangle $ $\leftrightarrow $ $|r\rangle $ transitions, and the
detuning $\delta \equiv \omega _{0}-\omega _{L}$ ($\omega _{L}=\omega
_{1}-\omega _{2}$) \cite{nist,dav,stein}. The ion interacts with an
effective laser plane wave propagating along the $x$ direction, with wave
vector $k_{L}=\omega _{L}/c$. In this configuration, only the ionic motion
along the $x$ axis will be modified. The transition between $|\downarrow
\rangle $ and a fourth level $\left| d\right\rangle $, achieved by another
laser strongly coupled to the electronic ground state, is analyzed in order
to measure the ionic vibrational state by collecting the resonance
fluorescence signal, which is the probability of the ion being found in the
internal state $\left| \downarrow \right\rangle $ \cite{meekhof}.

The ion-laser interaction Hamiltonian that describes the effective
interaction of the quantized motion of the ionic center of mass (CM) coupled
to its electronic degrees of freedom is \cite{nist,dav,meekhof} 
\begin{equation}
H_{ion-laser}=\hslash \Omega \left( \sigma _{+}%
%TCIMACRO{\limfunc{e}}%
%BeginExpansion
\mathop{\rm e}%
%EndExpansion
\nolimits^{ik_{L}x-i\omega _{L}t+i\phi }+\sigma _{-}%
%TCIMACRO{\limfunc{e}}%
%BeginExpansion
\mathop{\rm e}%
%EndExpansion
\nolimits^{-ik_{L}x+i\omega _{L}t-i\phi }\right) ,  \label{1}
\end{equation}%
where $\sigma _{+}=|\uparrow \rangle \langle \downarrow |$, $\sigma
_{-}=|\downarrow \rangle \langle \uparrow |$ and $\sigma _{z}$ are the usual
Pauli pseudo spin operators, $x$ is the position operator for the $x$
coordinate of the ion, $\Omega $ is the effective Rabi frequency of the
transition $\left| \uparrow \right\rangle \leftrightarrow \left| \downarrow
\right\rangle $, and $\phi $ is the phase difference between the two lasers.

The ion-BG interaction (the polarization of the BG induced by the
oscillation of the trapped ion) will be described by a Fr\"{o}hlich-type
electron-phonon Hamiltonian \cite{frohlich}. In Appendix A we show, by
analogy with surface electrons on liquid helium \cite%
{lifshitz,sabisky,shikin,studart}, that a polaron-like interaction results
when the electric field of the trapped ion polarizes the neutral BG (usually
H$_2$ in the NIST $^9$Be$^{+}$ experiments \cite{nist}). The attractive
ion-BG interaction potential is given by ${\rm U}(r)=-\chi q^2/\left( 8\pi
\epsilon _0r^4\right) $, where $\chi $ is the polarizability and $q$ the ion
charge, and the resulting Hamiltonian reads 
\begin{equation}
H_{ion-BG}=\sum_k\hslash V_k\left( b_k%
%TCIMACRO{\limfunc{e} }%
%BeginExpansion
\mathop{\rm e}
%EndExpansion
\nolimits^{ikx}+b_k^{\dagger }%
%TCIMACRO{\limfunc{e} }%
%BeginExpansion
\mathop{\rm e}
%EndExpansion
\nolimits^{-ikx}\right) ,  \label{2}
\end{equation}
where $b_k^{\dagger }$ $(b_k)$ is the creation (annihilation) operator of
BG-oscillation quanta, $V_k$ stands for the coupling strength, and $k$ are
the $x$-components of the BG-oscillation wave vector. In a frame rotating at
the ``effective laser frequency'' $\omega _L$, the ion-laser and ion-BG
Hamiltonians are given in the Schr\"{o}dinger picture by ($\hslash =1$ from
here on):

\begin{eqnarray}
H_{ion-laser} &=&\Omega \left( \sigma _{+}%
%TCIMACRO{\limfunc{e}}%
%BeginExpansion
\mathop{\rm e}%
%EndExpansion
\nolimits^{i\eta _{L}(a+a^{\dagger })-i\phi }+\sigma _{-}%
%TCIMACRO{\limfunc{e}}%
%BeginExpansion
\mathop{\rm e}%
%EndExpansion
\nolimits^{-i\eta _{L}(a+a^{\dagger })+i\phi }\right) ,  \label{3} \\
H_{ion-BG} &=&\sum_{k}V_{k}\left( b_{k}%
%TCIMACRO{\limfunc{e}}%
%BeginExpansion
\mathop{\rm e}%
%EndExpansion
\nolimits^{i\eta _{k}(a+a^{\dagger })}+b_{k}^{\dagger }%
%TCIMACRO{\limfunc{e}}%
%BeginExpansion
\mathop{\rm e}%
%EndExpansion
\nolimits^{-i\eta _{k}(a+a^{\dagger })}\right) ,  \label{4}
\end{eqnarray}%
where $a^{\dagger }$($a$) is the creation (annihilation) operator of
vibrational quanta, $\eta _{L}=$ $k_{L}/\sqrt{2m\nu }$ is the Lamb-Dicke
parameter, and $\eta _{k}=$ $k/\sqrt{2m\nu }$ stand for Lamb-Dicke-like
parameters due to the ion-BG interaction. The total Hamiltonian is given in
the Schr\"{o}dinger picture by $H=H_{0}+H_{ion-laser}+H_{ion-BG}$, where $%
H_{0}$ indicates the free Hamiltonian composed of the internal and motional
degrees of freedom of the trapped ion plus the BG: 
\begin{equation}
H_{0}=\nu a^{\dagger }a+\frac{\delta }{2}\sigma _{z}+\sum_{k}\omega
_{k}b_{k}^{\dagger }b_{k}.  \label{5}
\end{equation}%
Writing the total Hamiltonian in the interaction picture (bold labels), by
the unitary transformation $U(t)=\exp \left( -iH_{0}t\right) $, and then
expanding the resulting expressions in terms of the parameters $\eta _{L}$
and $\eta _{k}$, we get 
\begin{eqnarray}
{\bf H}_{ion-laser} &=&\Omega e^{-\eta _{L}^{2}/2}\left(
\sum_{m,l=0}^{\infty }\frac{(i\eta _{L})^{m+l}}{m!l!}\sigma _{+}a^{{\dagger }%
^{m}}a^{l}%
%TCIMACRO{\limfunc{e}}%
%BeginExpansion
\mathop{\rm e}%
%EndExpansion
\nolimits^{i\left[ (m-l)\nu +\delta \right] t-i\varphi }+h.c.\right) ,
\label{6} \\
{\bf H}_{ion-BG} &=&\sum_{k}V_{k}e^{-\eta _{k}^{2}/2}\left(
\sum_{m,l=0}^{\infty }\frac{(i\eta _{k})^{m+l}}{m!l!}b_{k}a^{{\dagger }%
^{m}}a^{l}%
%TCIMACRO{\limfunc{e}}%
%BeginExpansion
\mathop{\rm e}%
%EndExpansion
\nolimits^{i\left[ (m-l)\nu -\omega _{k}\right] t}+h.c.\right) ,  \label{7}
\end{eqnarray}%
where ion-laser resonance is achieved by tuning the laser frequencies to
obtain $\delta =-\ell \nu $ $(\ell =m-l)$. In what follows some reasonable
approximations are made, in order to simplify considerably the Hamiltonians (%
\ref{6}) and (\ref{7}). First we mention $(i)$ the standard Lamb-Dicke limit
for the ion-laser interaction, for which $\eta _{L}\ll 1$ {\bf (}where the
ionic CM motion is strongly localized with respect to the laser wavelengths%
{\bf )}. Next, from the low energy of the BG oscillations $(ii)$ a
Lamb-Dicke-like limit, $\eta _{k}\ll 1$, will be assumed for the ion-BG
interaction. This limit is somewhat analogous to the case of large polarons
in solid state physics \cite{devreese}. In fact, the motion of large
polarons is continuous, as should be that of a trapped ion. In contrast,
small polarons recognize the periodicity of a solid, becoming localized as
in the strong coupling theory and assuming atomic dimensions. With these two
approximations we obtain the simplified Hamiltonians 
\begin{eqnarray}
{\bf H}_{ion-laser} &=&\Omega \left( \sigma _{+}%
%TCIMACRO{\limfunc{e}}%
%BeginExpansion
\mathop{\rm e}%
%EndExpansion
\nolimits^{-i\delta t-i\varphi }+i\eta _{L}\sigma _{+}a^{{\dagger }}%
%TCIMACRO{\limfunc{e}}%
%BeginExpansion
\mathop{\rm e}%
%EndExpansion
\nolimits^{i\left( \nu +\delta \right) t-i\varphi }+i\eta _{L}\sigma _{+}a%
%TCIMACRO{\limfunc{e}}%
%BeginExpansion
\mathop{\rm e}%
%EndExpansion
\nolimits^{-i\left( \nu -\delta \right) t-i\varphi }+h.c.\right) ,  \label{8}
\\
{\bf H}_{ion-BG} &=&\sum_{k}V_{k}\left( b_{k}%
%TCIMACRO{\limfunc{e}}%
%BeginExpansion
\mathop{\rm e}%
%EndExpansion
\nolimits^{-i\omega _{k}t}+i\eta _{k}b_{k}a^{{\dagger }}%
%TCIMACRO{\limfunc{e}}%
%BeginExpansion
\mathop{\rm e}%
%EndExpansion
\nolimits^{i\left( \nu -\omega _{k}\right) t}+i\eta _{k}b_{k}a%
%TCIMACRO{\limfunc{e}}%
%BeginExpansion
\mathop{\rm e}%
%EndExpansion
\nolimits^{-i\left( \nu +\omega _{k}\right) t}+h.c.\right) .  \label{9}
\end{eqnarray}%
With the addition of $(iii)$ the optical rotating wave approximation, three
specific Hamiltonians for the ion-laser interaction are obtained, depending
on the choice of the ion-laser detuning:

(a) the carrier Hamiltonian ($\delta =0$),

\begin{equation}
{\bf H}_{ion-laser}^C=\Omega \left( \sigma _{+}e^{-i\varphi }+\sigma
_{-}e^{+i\varphi }\right) ,  \label{10}
\end{equation}
which induces the transition $\left| n,\downarrow \right\rangle
\longleftrightarrow \left| n,\uparrow \right\rangle $ (where $\left|
n\right\rangle $ indicates a motional Fock state), and is responsible for
rotating only the internal electronic levels of the ion wave function in
accordance with

\begin{mathletters}
\begin{eqnarray}
e^{-iH_{ion-laser}^C\tau }\left| n,\uparrow \right\rangle &=&\cos \left(
\Omega \tau \right) \left| n,\uparrow \right\rangle -ie^{i\varphi }\sin
\left( \Omega \tau \right) \left| n,\downarrow \right\rangle ,  \label{11a}
\\
e^{-iH_{ion-laser}^C\tau }\left| n,\downarrow \right\rangle &=&\cos \left(
\Omega \tau \right) \left| n,\downarrow \right\rangle -ie^{-i\varphi }\sin
\left( \Omega \tau \right) \left| n,\uparrow \right\rangle ;  \label{11b}
\end{eqnarray}

(b) the Jaynes-Cummings Hamiltonian ($\delta =\nu $), corresponding to the
first red sideband, 
\end{mathletters}
\begin{equation}
{\bf H}_{ion-laser}^{JC}=i\eta _L\Omega \left( \sigma _{+}ae^{-i\varphi
}-\sigma _{-}a^{\dagger }e^{+i\varphi }\right) ,  \label{12}
\end{equation}
induces the transition $\left| n,\downarrow \right\rangle
\longleftrightarrow \left| n-1,\uparrow \right\rangle $, in such a way that
the electronic and vibrational\ modes evolve as

\begin{mathletters}
\begin{eqnarray}
e^{-iH_{ion-laser}^{JC}\tau }\left| n,\uparrow \right\rangle &=&C_n\left|
n,\uparrow \right\rangle -e^{-i\varphi }S_n\left| n+1,\downarrow
\right\rangle ,  \label{13a} \\
e^{-iH_{ion-laser}^{JC}\tau }\left| n,\downarrow \right\rangle
&=&C_{n-1}\left| n,\downarrow \right\rangle +e^{i\varphi }S_{n-1}\left|
n-1,\uparrow \right\rangle ,  \label{13b}
\end{eqnarray}
where $C_n=\cos (g\tau \sqrt{n+1})$, $S_n=\sin (g\tau \sqrt{n+1})$, $\tau $
is the duration of the laser pulses, and $g=\eta _L\Omega $;

(c) the Anti-Jaynes-Cummings Hamiltonian ($\delta =-\nu $), corresponding to
the first blue sideband, 
\end{mathletters}
\begin{equation}
{\bf H}_{ion-laser}^{AJC}=i\eta _L\Omega \left[ \sigma _{+}a^{{\dagger }%
}e^{-i\varphi }-\sigma _{-}ae^{+i\varphi }\right] ,  \label{14}
\end{equation}
induces the transition $\left| n,\downarrow \right\rangle
\longleftrightarrow \left| n+1,\uparrow \right\rangle $, and the electronic
and vibrational\ modes evolve as

\begin{mathletters}
\begin{eqnarray}
e^{-iH_{ion-laser}^{AJC}\tau }\left| n,\downarrow \right\rangle
&=&C_{n}\left| n,\downarrow \right\rangle +e^{-i\varphi }S_{n}\left|
n+1,\uparrow \right\rangle ,  \label{140a} \\
e^{-iH_{ion-laser}^{AJC}\tau }\left| n,\uparrow \right\rangle
&=&C_{n-1}\left| n,\uparrow \right\rangle -e^{i\varphi }S_{n-1}\left|
n-1,\downarrow \right\rangle .  \label{140b}
\end{eqnarray}%
Finally, assuming hypothesis $(iii)$ and that $(iv)${\it \ }the
BG-oscillation modes are closely spaced in frequency, with the maximum of
their spectrum far from zero, the ion-BG Hamiltonian becomes

\end{mathletters}
\begin{equation}
{\bf H}_{ion-BG}=i\sum_k\eta _kV_k\left( b_ka^{{\dagger }}%
%TCIMACRO{\limfunc{e} }%
%BeginExpansion
\mathop{\rm e}
%EndExpansion
\nolimits^{i(\nu -\omega _k)}-b_k^{\dagger }a%
%TCIMACRO{\limfunc{e} }%
%BeginExpansion
\mathop{\rm e}
%EndExpansion
\nolimits^{-i(\nu -\omega _k)}\right) ,  \label{15}
\end{equation}
which plays the role of a reservoir for the motional degree of freedom of
the trapped ion.

Next, we analyze the effects of the ion-BG coupling on the time evolution of
the ionic external and internal degrees of freedom. We note that the
ion-laser detunings $\delta =\pm \nu $ cause the internal levels of the
trapped ion to be affected by the BG, through the motional degree of
freedom, differently from the choice $\delta =0$, which leaves the BG
without effect on the internal levels of the trapped ion.

\section{Effects of the ion-BG coupling on a prepared motional-electronic
state}

In what follows, the effects of the ion-BG coupling on the behavior of a
prepared motional-electronic state of the trapped ion will be considered for
the ion-laser detunings $\delta =0$ and $-\nu $, representing the Carrier
and the Anti-Jaynes-Cummings ion-laser interactions discussed above.

\subsection{Carrier Hamiltonian}

First, using Glauber's $P$ representation \cite{glauber} we briefly analyze
the dynamics of the ionic motional state when $\delta =0$, the total
Hamiltonian of the system, in the Schr\"{o}dinger picture, being 
\begin{eqnarray}
H &=&H_{0}+U^{\dagger }(t){\bf H}_{ion-laser}^{C}U(t)+U^{\dagger }(t){\bf H}%
_{ion-BG}U(t)  \nonumber \\
&=&\nu a^{\dagger }a+\sum_{k}\omega _{k}b_{k}^{\dagger }b_{k}+\Omega \left(
\sigma _{+}e^{-i\varphi }+\sigma _{-}e^{+i\varphi }\right) +i\sum_{k}\eta
_{k}V_{k}\left( b_{k}a^{{\dagger }}-b_{k}^{\dagger }a\right)  \label{16}
\end{eqnarray}%
Focusing on the time evolution of the ionic motional state, which is not
affected by the internal levels in the Carrier regime, we observe that the
Hamiltonian (\ref{16}) is analogous to that of a single-mode field trapped
in a lossy cavity. The latter Hamiltonian has received considerable
attention recently for computing the fidelity of a cavity-field state
required for quantum communication or computation purposes \cite{ANM}. In
this connection, we find it interesting to analyze the decoherence process
of a prepared motional state of a trapped ion due to its coupling to the BG.

Solving the coupled linear equations of motion for the operators $a$ and $%
b_{k}$, we obtain from Eq. (\ref{16}) 
\begin{equation}
a(t)=u(t)a(0)+\sum_{k}v_{k}(t)b_{k}(0),  \label{17}
\end{equation}%
where, disregarding the typical small frequency shifts and introducing the
damping constant $\Gamma $, the time-dependent coefficients, obeying the
initial conditions $u(0)=1$ and $v_{k}(0)=0$, can be written as 
\begin{mathletters}
\begin{eqnarray}
u(t) &=&%
%TCIMACRO{\limfunc{e}}%
%BeginExpansion
\mathop{\rm e}%
%EndExpansion
\nolimits^{-(\Gamma /2+i\nu )t},  \label{18a} \\
v_{k}(t) &=&-\eta _{k}V_{k}%
%TCIMACRO{\limfunc{e}}%
%BeginExpansion
\mathop{\rm e}%
%EndExpansion
\nolimits^{-i\omega _{k}t}\frac{1-%
%TCIMACRO{\limfunc{e}}%
%BeginExpansion
\mathop{\rm e}%
%EndExpansion
{}^{i(\omega _{k}-\nu )t}%
%TCIMACRO{\limfunc{e}}%
%BeginExpansion
\mathop{\rm e}%
%EndExpansion
{}^{-\Gamma t/2}}{\Gamma /2-i(\omega _{k}-\nu )}.  \label{18b}
\end{eqnarray}

From the above results, we obtain the normally ordered characteristic
function $\chi _{N}(\xi ,t)$, defined in the Schr\"{o}dinger picture as the
expectation value 
\end{mathletters}
\begin{equation}
\chi _{N}(\xi ,t)={\rm {Tr}}\left[ \rho (t)%
%TCIMACRO{\limfunc{e}}%
%BeginExpansion
\mathop{\rm e}%
%EndExpansion
\nolimits^{\xi a^{\dagger }(0)}%
%TCIMACRO{\limfunc{e}}%
%BeginExpansion
\mathop{\rm e}%
%EndExpansion
\nolimits^{\xi ^{\ast }a(0)}\right] ,  \label{19}
\end{equation}%
where the density operator $\rho (t)$ refers to the composite ion-BG system,
assumed, as usual, to be initially decoupled, $\rho (0)=\rho _{ion}(0)\rho
_{BG}(0)$ (the ion-BG coupling is turned on suddenly at $t=0^{+}$). For a
motional state of the trapped ion, prepared as $\sum_{l}c_{l}\left| \alpha
_{l}\right\rangle $, follows the representative term of the reduced density
operator $\rho _{ion}(0)=\left| \alpha _{1}\right\rangle \left\langle \alpha
_{2}\right| $, while for an initially thermal state with Gaussian
distribution the reduced density operator for the BG reads 
\begin{equation}
\rho _{BG}(0)=\prod_{k}\int \frac{%
%TCIMACRO{\limfunc{e}}%
%BeginExpansion
\mathop{\rm e}%
%EndExpansion
{}^{-\left| \beta _{k}\right| ^{2}/\left\langle n_{k}\right\rangle }}{\pi
\left\langle n_{k}\right\rangle }\left| \beta _{k}\right\rangle \left\langle
\beta _{k}\right| {\rm {d}}^{2}\beta _{k}.  \label{20}
\end{equation}

From the characteristic function (\ref{19}) we derive the conditional
distribution function 
\begin{eqnarray}
P(\gamma ,t) &=&\frac{1}{\pi ^{2}}\int 
%TCIMACRO{\limfunc{e}}%
%BeginExpansion
\mathop{\rm e}%
%EndExpansion
\nolimits^{\gamma \xi ^{\ast }-\gamma ^{\ast }\xi }\chi _{N}(\xi ,t){\rm {d}}%
^{2}\xi  \nonumber \\
&=&\frac{\left\langle \alpha _{1}|\alpha _{2}\right\rangle }{\pi D(t)}\exp %
\left[ -\frac{\left( \gamma ^{\ast }-u^{\ast }(t)\alpha _{1}^{\ast }\right)
\left( \gamma -u(t)\alpha _{2}\right) }{D(t)}\right] ,  \label{21}
\end{eqnarray}%
whose dispersion is given by $D(t)=\sum_{k}\left\langle n_{k}\right\rangle
\left| v_{k}(t)\right| ^{2}=\overline{n}_{th}\left( 1-%
%TCIMACRO{\limfunc{e}}%
%BeginExpansion
\mathop{\rm e}%
%EndExpansion
\nolimits^{-\Gamma t}\right) $, where the occupation number of the
BG-oscillations in thermal equilibrium reads $\overline{n}_{th}=\left( 
%TCIMACRO{\limfunc{e}}%
%BeginExpansion
\mathop{\rm e}%
%EndExpansion
\nolimits^{-\nu /k_{B}T}-1\right) ^{-1}$, $k_{B}$ is the Boltzmann constant,
and $T$ \ is the absolute temperature. Fig. 2 display the damping process in
the evolution of the conditional distribution function $P(\gamma ,t)$ for a
prepared motional coherent state. The rotation in phase space arises from
the factor $\exp (-i\nu t)$ in Eq. (\ref{18a}) and we observe both the
effects coming from the environment: the loss of excitation, carrying the
initial coherent state to the vacuum state, and the diffusion due to the
nonzero temperature of the BG. These effects are clearly seen from the shape
of the distribution function $P(\gamma ,t)$ depicted in Fig. 2 at $\Gamma
t=0.2$ and $0.9$.

From the distribution function $P(\gamma ,t)$, we can obtain the reduced
density operator for the ion, 
\begin{equation}
\rho _{ion}(t)=\frac{1}{\pi ^{2}}\int P(\gamma ,t)\left| \gamma
\right\rangle \left\langle \gamma \right| d^{2}\gamma ,  \label{210}
\end{equation}%
and the mean value of operators associated with the ionic motional states;
for example, the mean excitation of a prepared motional vacuum state, i.e., $%
\alpha _{1}=\alpha _{2}=0$, reads 
\begin{equation}
\left\langle a^{\dagger }(t)a(t)\right\rangle =\int P(\alpha ,t)\alpha
^{\ast }\alpha d^{2}\alpha =D(t),  \label{23}
\end{equation}%
showing the heating process of the motional vacuum state due to a thermal BG.

\subsection{Anti-Jaynes-Cummings Hamiltonian}

Next, we examine the action of the ion-BG coupling on a prepared
motional-electronic state of the trapped ion when $\delta =-\nu $, the total
Hamiltonian being 
\begin{eqnarray}
H &=&H_{0}+U^{\dagger }(t){\bf H}_{ion-laser}^{AJC}U(t)+U^{\dagger }(t){\bf H%
}_{ion-BG}U(t)  \nonumber \\
&=&\nu a^{\dagger }a-\frac{\nu }{2}\sigma _{z}+\sum_{k}\omega
_{k}b_{k}^{\dagger }b_{k}+i\eta _{L}\Omega \left( \sigma _{+}a^{{\dagger }%
}e^{-i\varphi }-\sigma _{-}ae^{+i\varphi }\right) +i\sum_{k}\eta
_{k}V_{k}\left( b_{k}a^{{\dagger }}-b_{k}^{\dagger }a\right) .  \label{24}
\end{eqnarray}%
We are considering the Anti-Jaynes-Cummings (instead of the Jaynes-Cummings)
Hamiltonian for the ion-laser interaction, so as to compute, in this regime,
the damping of the Rabi oscillations of a trapped ion initially cooled to
its motional ground state and to compare our results with the available
experimental data for a trapped $^{9}$Be$^{+}$ ion \cite{gatoW2}.

To compute the atomic inversion from Hamiltonian (\ref{24}), we use the
techniques developed to investigate the Jaynes-Cummings Hamiltonian for an
atom interacting with a monochromatic field trapped in a lossy cavity at a
finite temperature, which is analogous to Eq. (\ref{24}) and has received
considerable attention in the literature \cite%
{sachdev,barnett,agarwal,scully}. The standard master-equation technique was
employed in Ref. \cite{barnett} to obtain the density matrix for the
combined atom-field system in the interaction picture. However, following %
\cite{sachdev}, we calculate the evolution of the atomic inversion from the
time-dependent averages of atomic excitation, photon numbers, and other
field quantities. Although also based on the master-equation approach, the
technique in \cite{sachdev} consists of working with a hierarchy of $c$%
-number quantities derived from the equation of motion for any operator of
the form $\left( a^{\dagger }\right) ^{m}a^{n}{\cal O}_{A}$ (where ${\cal O}%
_{A}$ is an atomic operator), given by 
\begin{eqnarray}
\frac{d}{dt}\left[ \left( a^{\dagger }\right) ^{m}a^{n}{\cal O}_{A}\right]
&=&-i\left[ \left( a^{\dagger }\right) ^{m}a^{n}{\cal O}_{A},-\frac{\nu }{2}%
\sigma _{z}+i\eta _{L}\Omega \left( \sigma _{+}a^{{\dagger }}e^{-i\varphi
}-\sigma _{-}ae^{+i\varphi }\right) \right]  \nonumber \\
&&+\left\langle \frac{d}{dt}\left[ \left( a^{\dagger }\right) ^{m}a^{n}%
\right] \right\rangle _{BG}{\cal O}_{A},  \label{25}
\end{eqnarray}%
where the last term includes the commutation relation between $\left(
a^{\dagger }\right) ^{m}a^{n}{\cal O}_{A}$ and the components of the
Hamiltonian (\ref{24}) describing the energy of the BG and its interaction
with the ionic external modes, as well as the motional energy of the trapped
ion. Under the Wigner-Weisskopf (WW) approximation and supposing that the BG
is in thermal equilibrium, we obtain \cite{scully} 
\begin{eqnarray}
\left\langle \frac{d}{dt}\left[ \left( a^{\dagger }\right) ^{m}a^{n}\right]
\right\rangle _{BG} &=&\left( i\nu (m-n)-\frac{\Gamma }{2}(m+n)\right)
\left\langle \left( a^{\dagger }\right) ^{m}a^{n}\right\rangle _{BG} 
\nonumber \\
&&+\Gamma mn\overline{n}_{th}\left\langle \left( a^{\dagger }\right)
^{m-1}a^{n-1}\right\rangle _{BG}.  \label{26}
\end{eqnarray}%
The damping constant, arising from the WW approximation, is $\Gamma =2\pi %
\left[ g(\nu )\right] ^{2}\Lambda (\nu )$, where $g(\nu )\equiv g_{\nu /c}$
is the coupling constant evaluated at $k=\nu /c$ and $\Lambda (\nu )={\Bbb V}%
\nu ^{2}/\pi c^{3}$ (${\Bbb V}$ being the quantization volume) is the
density of states. The occupation number $\overline{n}_{th}$ follows from
the assumption of thermal equilibrium, where the noise operators ${\cal N}%
(t)=\sum_{k}g_{k}b_{k}(0)%
%TCIMACRO{\limfunc{e}}%
%BeginExpansion
\mathop{\rm e}%
%EndExpansion
\nolimits^{-i(\omega _{k}-\nu )t}$ (with $g_{k}=\eta _{k}V_{k}$) satisfy $%
\left\langle {\cal N}(t)\right\rangle _{BG}=\left\langle {\cal N}^{\dagger
}(t)\right\rangle _{BG}=0$, and $\left\langle {\cal N}^{\dagger }(t){\cal N}%
(t^{\prime })\right\rangle _{BG}=\Gamma \overline{n}_{th}\delta (t-t^{\prime
})$ (see Ref. \cite{scully}).

From Eqs. (\ref{25}) and (\ref{26}) we derive the time-evolution of the
atomic inversion $\left\langle \sigma _{z}\right\rangle $ and the motional
occupation number of the trapped ion $\left\langle a^{\dagger
}a\right\rangle :$ 
\begin{eqnarray}
\frac{d\left\langle \sigma _{z}\right\rangle }{dt} &=&2g\left\langle \sigma
_{+}a^{{\dagger }}e^{-i\varphi }+\sigma _{-}ae^{+i\varphi }\right\rangle ,
\label{27} \\
\frac{d\left\langle a^{\dagger }a\right\rangle }{dt} &=&g\left\langle \sigma
_{+}a^{{\dagger }}e^{-i\varphi }+\sigma _{-}ae^{+i\varphi }\right\rangle
-\Gamma \left\langle a^{\dagger }a\right\rangle +\Gamma \overline{n}_{th}
\label{28}
\end{eqnarray}%
with the angular brackets denoting the BG as well as the quantum mechanical
average. Eqs. (\ref{27}) and (\ref{28}) involve the average of the Hermitian
operator $\sigma _{+}a^{{\dagger }}e^{-i\varphi }+\sigma _{-}ae^{+i\varphi }$
whose equation of motion involves the quantity $\left\langle a^{\dagger
}a\sigma _{z}\right\rangle $. In its turn, the equation of motion of this
quantity involves higher-order products of averaged operators such as $%
\left\langle \left( a^{\dagger }\right) ^{2}a^{2}\right\rangle $, $%
\left\langle \left( a^{\dagger }\right) ^{2}a^{2}\sigma _{z}\right\rangle $,
and so on. To deal with such averaged operators it is convenient to define
the $c$-numbers variables 
\begin{mathletters}
\begin{eqnarray}
{\cal P}_{n} &=&\left\langle \left( a^{\dagger }\right)
^{n}a^{n}\right\rangle ,\quad n\geq 0,\quad {\cal P}_{0}=1;  \label{29a} \\
{\cal Q}_{n} &=&\left\langle \left( a^{\dagger }\right) ^{n}a^{n}\sigma
_{z}\right\rangle ,\quad n\geq 0,\quad {\cal Q}_{0}=\left\langle \sigma
_{z}\right\rangle ;  \label{29b} \\
{\cal R}_{n} &=&\left\langle \sigma _{+}\left( a^{{\dagger }}\right)
^{n}a^{n-1}e^{-i\varphi }+\sigma _{-}\left( a^{{\dagger }}\right)
^{n-1}a^{n}e^{+i\varphi }\right\rangle ,\quad n>0.  \label{29c}
\end{eqnarray}%
Solving the dynamical equations for the above $c$-number variables we obtain
the dynamics of the atomic inversion and the motional occupation number of
the trapped ion. From definitions (\ref{29a})-(\ref{29c}), and Eqs. (\ref{25}%
) and (\ref{26}) we obtain the following set of equations 
\end{mathletters}
\begin{mathletters}
\begin{eqnarray}
\frac{d{\cal P}_{n}}{dt} &=&ng{\cal R}_{n}-n\Gamma {\cal P}_{n}+n^{2}\Gamma 
\overline{n}_{th}{\cal P}_{n-1},  \label{30a} \\
\frac{d{\cal Q}_{n}}{dt} &=&ng{\cal R}_{n}+2g{\cal R}_{n+1}-n\Gamma {\cal Q}%
_{n}+n^{2}\Gamma \overline{n}_{th}{\cal Q}_{n-1},  \label{30b} \\
\frac{d{\cal R}_{n}}{dt} &=&-2g{\cal Q}_{n}+ng{\cal P}_{n-1}-ng{\cal Q}%
_{n-1}-(n-1/2)\Gamma {\cal R}_{n}+n(n-1)\Gamma \overline{n}_{th}{\cal R}%
_{n-1}.  \label{30c}
\end{eqnarray}%
Note that at zero temperature ($\overline{n}_{th}=0$) and an initially
prepared ionic state $\left| n=0,\downarrow \right\rangle $ we obtain a
closed set of equations, since the expectation value of the operators
involving quadratic or higher powers of the ionic motional operators $a$ and 
$a^{\dagger }$ are therefore zero at all times \cite{scully}. However, to
compare the decay of Rabi oscillations resulting from our model with the
experimental data \cite{gatoW2}, we have to deal with the realistic case of
non-zero temperature. Observing that for the initially prepared ionic state $%
\left| n=0,\downarrow \right\rangle $, the Anti-Jaynes-Cummings Hamiltonian
induces the transition $\left| 0,\downarrow \right\rangle
\longleftrightarrow \left| 1,\uparrow \right\rangle $, and it is possible to
estimate a temperature around $\nu /k_{B}$ for the trapped ion and,
consequently, for the BG. In fact, under such conditions the thermal energy
of the trapped ion $k_{B}T$ is around $\nu $, resulting in a motional
occupation number around unity ($\overline{n}_{th}\approx 1$). On the other
hand, since the Rabi oscillations in NIST experiments \cite{gatoW2} survive
considerably above $100\mu s$, it is possible to estimate a damping factor $%
\Gamma $ around $10^{-3}s$. Having these values at hand and truncating the
system of equations (\ref{30a})-(\ref{30c}) at $n=4$ (an approximation due
to the low motional occupation number estimated above for the BG), we
obtain, via the numerical Laplace transform method, the time-evolution of
the probability for fluorescence measurement of the electronic ground state $%
{\sf P}_{\downarrow }(t)=\left( 1-\left\langle \sigma _{z}\right\rangle
\right) /2$. Fig. 3 displays the behavior of the function ${\sf P}%
_{\downarrow }(t)$ computed from our model, which is in excellent agreement
with the experimental data reported in \cite{gatoW2}. We used in our
numerical calculation the experimental parameters $\eta _{L}=0.202$ and $%
\Omega /2\pi \approx 475$ kHz. In particular, our model reproduces the
asymmetry of the decay of Rabi oscillations seen in the experimental data.
To obtain such behavior, we used the parameters $\overline{n}_{th}=1.0$ and $%
\Gamma /g=$ $6.0\times $\thinspace $10^{-3}$, which are in good agreement
with the values estimated above. It is worth noting that, when switching off
the laser pulse, Eq. (\ref{23}) indicates that the motional occupation
number of the trapped ion $\left\langle a^{\dagger }a\right\rangle $ goes
asymptotically to unity, as expected.

In order to demonstrate from our model the sensitivity of the damping of
Rabi oscillations to the motional quantum number, we depicted in Figs.
4(a,b) the behavior of the function ${\sf P}_{\downarrow }(t)$ computed from
our model (full lines) when the initially prepared ionic states are $\left|
n=0,\downarrow \right\rangle $ [Fig. 4(a)] and $\left| n=1,\downarrow
\right\rangle $ [Fig. 4(b)]. Figs. 4(a,b) also display the heuristic
relation used in \cite{gatoW2} for fitting the experimental data, 
\end{mathletters}
\begin{equation}
{\sf P}_{\downarrow }(t)\approx \frac{1}{2}\left( 1+\sum_{n}p_{n}\cos
(2\Omega t\sqrt{n+1})e^{-\gamma _{n}t}\right) ,  \label{31}
\end{equation}%
where $p_{n}$ is the initial probability distribution of the motional states
in the Fock basis and $\gamma _{n}=\gamma _{0}(n+1)^{0.7}$ is a
phenomenological damping rate. The behavior of the heuristic relation (\ref%
{31}) for $\left| n=0,\downarrow \right\rangle $ (Fig. 4(a)) and $\left|
n=1,\downarrow \right\rangle $ (Fig. 4(b)) is shown in dotted lines using
the value $\gamma _{0}=11.9(4)$ kHz estimated in \cite{gatoW2}. We stress
that Eq. (\ref{31}) describes a symmetric decay of the Rabi oscillations
instead of the experimentally observed asymmetry which is in agreement with
our model (see Fig. 3). Therefore, we claim that the polarization of the
residual background gas is the mechanism leading to the $n$-dependence of
the decoherence effects of a trapped ion.

\section{Comments and Conclusions}

We have proposed an alternative mechanism to explain the damping of Rabi
oscillations of a $^{9}$Be$^{+}$ ion initially cooled to its motional ground
state. Our model allows for the influence of the polarization of the
residual background gas induced by the oscillating trapped ion. The
polarization of the BG, which in turn influences the trapped ion motion, is
described by a Fr\"{o}hlich-type ion-BG interaction with a Lamb-Dicke-like
limit where $\eta _{k}=k/\sqrt{2m\nu }\ll 1$, $k$ being the BG oscillation
wave vectors, and $m$ ($\nu $) being the mass (frequency) of the trapped ion.

Since experiments on trapped ions are carried out in a high-vacuum
environment, the density of the BG, around $10^6$ cm$^{-3}$ \cite{nist},
does not permit a polaron binding energy and the trapped ion is scattered by
the BG. Both elastic and inelastic collisions are present, although elastic
collisions are expected to be the main source of decoherence. The
troublesome inelastic processes of chemical reactions and charge exchange
are expected to be rare. Although a conservative estimate for the elastic
collisions rate also leads to a small value, we claim that the damping
medium provided by the background, which allows laser-cooling to proceed
faster when ions are first loaded into a trap, has also to be taken into
account when the trapped ions are cooled to their motional ground state. As
mentioned in \cite{nist}, for a linear Paul-rf trap the main source of
decoherence may be associated with errors arising from the external degrees
of freedom of the trapped ion rather than internal levels or a nonideal
applied field.

Under the Lamb-Dicke-like limit adopted here, the resulting ion-BG
interaction consisting of the standard ``rotating wave'' terms (assuming
that the spectrum of the phonon modes presents its maximum far from zero)
turns out to be the usual linear response model of the reservoir (the BG) to
the system (the trapped ion) \cite{glauber,caldeira}. Glauber's $P$
representation was used to analyze the dynamics of the ionic motional state
when considering a Carrier pulse for the ion-laser interaction. The Carrier
pulse prevents the BG having any effect on the internal levels of the
trapped ion and either the damping or the heating process of the ionic
motional state, depending on the temperature of the BG, was investigated. On
the other hand, a master-equation approach was used to achieve a theoretical
fit to the exponentially decaying, sinusoidal Rabi oscillations \cite{gatoW2}
resulting from a prepared motional-electronic state of the trapped ion under
the anti-Jaynes-Cummings pulse.

The complete agreement between the measured data for Rabi oscillations with
the behavior computed from our model indicates that polarization effects
constitute a relevant source of error in the ionic trap. Evidently, the
error from polarization of the BG has to be taken together with the errors
coming from the internal levels and the application of a nonideal field \cite%
{nist}, as investigated in Refs. \cite{milburn,vogeln}. Like the stochastic
approach in \cite{vogeln}, our model also reproduces the asymmetry of the
decay of Rabi oscillations, as observed in the experimental data. Since the
authors in Ref. \cite{vogeln} claim that the asymmetry of the measured data
is consistent with the assumption of the mechanism of quantum jumps, we
mention that it is also consistent with the damping-heating competition
produced by the ion-BG interaction (through the motional occupation number $%
\overline{n}_{th}$ and the damping factor $\Gamma $). Moreover, the
consistency of our model is confirmed by the fact that the best fit of the
measured data was accomplished with parameter values ($\overline{n}_{th}=1.0$
and $\Gamma /g=$ $6.0\times $\thinspace $10^{-3}$) in agreement with
theoretical estimates.

Finally, we have demonstrated with our model that the damping of Rabi
oscillations is sensitive to the motional quantum number, as experimentally
observed. The behaviour of the curves in Figs. 4(a,b), representing two
ionic states initially prepared with different motional quantum numbers $%
\left| n=0,\downarrow \right\rangle $ and $\left| n=1,\downarrow
\right\rangle $, respectively, strongly suggests that the polarization of
the residual background gas is the mechanism leading to the $n$-dependence
of the decoherence effects of a trapped ion.

\bigskip

\begin{center}
{\bf Acknowledgments}
\end{center}

We wish to express thanks for the support from CNPq and FAPESP, Brazilian
agencies. We also thank S. S. Sokolov and N. Studart for helpful discussions
and D. J. Wineland and C. Monroe for sending us the experimental data used
in Fig. 3.

{\bf \ }{\LARGE Appendix A}

In this appendix, reasoning by analogy with the physics of surface electrons
on liquid helium \cite{shikin,studart}, we show how to obtain a Fr\"{o}%
hlich-type ion-BG interaction. For practical reason, we assume a continuous
BG and a cylindrical symmetry for the ion-BG system, as depicted in Fig. 5.
The ion, oscillating along the $x$-axis, is assumed to be at a distance $z$
from the origin of the coordinate system which is fixed in the center of the
BG surface located in the $xy$-plane. This distance $z$ (which contributes
to the coupling parameter $V_{k}$) is here assumed to be a mean distance
between the ion and the atoms composing the BG, a mean interparticle spacing
between two colliding partners. As mentioned in the Introduction, the ion is
scattered by the oscillations of the BG surface, whose dynamic roughness is
given by $\xi \left( \overrightarrow{r^{\prime }}\right) $ (see Fig. 5).

The potential energy of the ion due to polarization of the BG is

\[
{\cal U}\left( \overrightarrow{r},z\right) =-\frac{\chi q^2}2\int d^2%
\overrightarrow{r^{\prime }}\int_{-\infty }^{\xi \left( \overrightarrow{%
r^{\prime }}\right) }dz^{\prime }\frac 1{\left[ \left| \overrightarrow{%
r^{\prime }}-\overrightarrow{r}\right| ^2+\left( z^{\prime }-z\right) ^2%
\right] ^2}. 
\]
Substituting the variables $z-\xi \left( \overrightarrow{r^{\prime }}\right)
=\zeta $, $z^{\prime }-\xi \left( \overrightarrow{r^{\prime }}\right) =\zeta
^{\prime }$, and defining $\Lambda /\pi \equiv \chi \rho e^2/2$, we can
rewrite the potential energy as

\begin{eqnarray*}
{\cal U}\left( \overrightarrow{r},z\right) &=&\delta {\cal U}\left( 
\overrightarrow{r},z\right) +{\cal U}_{0}\left( \overrightarrow{r},z\right) ,
\\
\delta {\cal U}\left( \overrightarrow{r},z\right) &=&-\frac{\Lambda }{\pi }%
\int d^{2}\overrightarrow{r^{\prime }}\int_{0}^{\xi \left( \overrightarrow{%
r^{\prime }}\right) -\xi \left( \overrightarrow{r}\right) }d\zeta ^{\prime }%
\frac{1}{\left[ \left| \overrightarrow{r^{\prime }}-\overrightarrow{r}%
\right| ^{2}+\left( \zeta ^{\prime }-\zeta \right) ^{2}\right] ^{2}}, \\
{\cal U}_{0}\left( \overrightarrow{r},z\right) &=&-\frac{\Lambda }{\pi }\int
d^{2}\overrightarrow{r\prime }\int_{\infty }^{0}d\zeta ^{\prime }\frac{1}{%
\left[ \left| \overrightarrow{r^{\prime }}-\overrightarrow{r}\right|
^{2}+\left( \zeta ^{\prime }-\zeta \right) ^{2}\right] ^{2}}.
\end{eqnarray*}%
Assuming that the roughness of the BG, $\xi \left( \overrightarrow{r^{\prime
}}\right) -\xi \left( \overrightarrow{r}\right) $, is sufficiently small and
expanding $\xi \left( \overrightarrow{r^{\prime }}\right) $ and $\xi \left( 
\overrightarrow{r}\right) $ in Fourier series we obtain

\begin{eqnarray*}
\delta {\cal U}\left( \overrightarrow{r},z\right) &\approx &-\frac \Lambda %
\pi \int d^2\overrightarrow{r^{\prime }}\left. \frac{\xi \left( 
\overrightarrow{r^{\prime }}\right) -\xi \left( \overrightarrow{r}\right) }{%
\left[ \left| \overrightarrow{r^{\prime }}-\overrightarrow{r}\right|
^2+\left( z^{\prime }-z\right) ^2\right] ^2}\right| _{z^{\prime }=0} \\
&=&\frac \Lambda {\pi \sqrt{S}}\sum_{\overrightarrow{k}}\xi \left( 
\overrightarrow{k}\right) e^{i\overrightarrow{k}\cdot \overrightarrow{r}}%
\left[ \int \frac{d^2\overrightarrow{\lambda }}{\left( \lambda ^2+z^2\right)
^2}-\int \frac{d^2\overrightarrow{\lambda }e^{i\overrightarrow{k}\cdot 
\overrightarrow{\lambda }}}{\left( \lambda ^2+z^2\right) ^2}\right] ,
\end{eqnarray*}
where $S$ is the area of the BG surface and we have substituted the variable 
$\overrightarrow{r^{\prime }}-\overrightarrow{r}=\overrightarrow{\lambda }$.
After integrating over $\overrightarrow{\lambda }$ we obtain the perturbed
potential energy

\[
\delta {\cal U}\left( \overrightarrow{r},z\right) \approx \sum_{%
\overrightarrow{k}}\xi \left( \overrightarrow{k}\right) e^{i\overrightarrow{k%
}\cdot \overrightarrow{r}}\frac{\Lambda }{z\sqrt{S}}\left[ \frac{1}{z}-k{\rm %
K}_{1}\left( kz\right) \right] , 
\]%
where ${\rm K}_{1}\left( kz\right) $ is the modified Bessel function. At
this point we can quantize the dynamic variable $\xi \left( \overrightarrow{k%
}\right) $ as follows \cite{shikin}:

\begin{eqnarray*}
\xi \left( \overrightarrow{k}\right) &=&C_{\overrightarrow{k}}\left( a_{%
\overrightarrow{k}}+a_{-\overrightarrow{k}}^{\dagger }\right) , \\
C_{\overrightarrow{k}} &=&\sqrt{\frac \hbar {2\rho \omega \left( 
\overrightarrow{k}\right) }}
\end{eqnarray*}
where $\omega \left( \overrightarrow{k}\right) $ is the dispersion relation.
Defining the coupling parameter

\[
V_{\overrightarrow{k}}=\frac{\Lambda C_{\overrightarrow{k}}}{z\sqrt{S}}\left[
\frac 1z-k{\rm K}_1\left( kz\right) \right] , 
\]
we finally obtain the ionic quantized potential energy due to interaction
with the BG as

\[
{\cal U}\left( \overrightarrow{r},z\right) =\sum_{\overrightarrow{k}}V_{%
\overrightarrow{k}}\left( a_{\overrightarrow{k}}+a_{-\overrightarrow{k}%
}^{\dagger }\right) e^{i\overrightarrow{k}\cdot \overrightarrow{r}}+{\cal U}%
_0\left( \overrightarrow{r},z\right) , 
\]
which results, in one dimension, in the interaction Hamiltonian (\ref{2}). $%
{\cal U}_0\left( \overrightarrow{r},z\right) $ plays the role of a reference
energy.

%%%%%%%%%%%%%%%%%%%%%%%%%%%%%%%%%%%%%%%%%%%%%%%%%%%%%%%%%%%%%%%%%%%%%%%
%

{\bf Figure Captions}

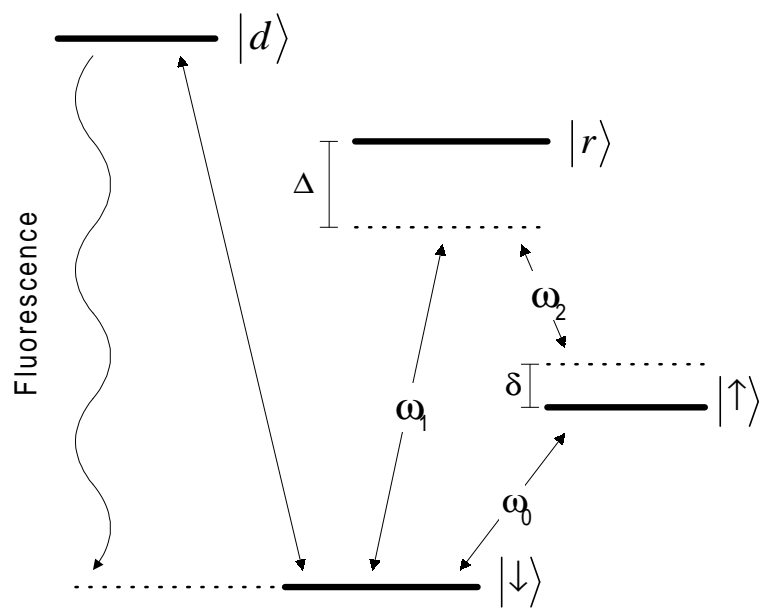
FIG. 1. \ Electronic energy level diagram of a trapped ion interacting with
laser beams of frequency $\omega _{1}$ and $\omega _{2}$, where $\delta
=\omega _{0}-\omega _{1}+\omega _{2}$ ($\delta \ll \Delta $), $\left|
r\right\rangle $ \ (adiabatically eliminated) is an auxiliary electronic
level which indirectly couples the levels $\left| \uparrow \right\rangle $
and $\left| \downarrow \right\rangle $, and $\left| d\right\rangle $ is an
electronic level used to measure the fluorescence emission.

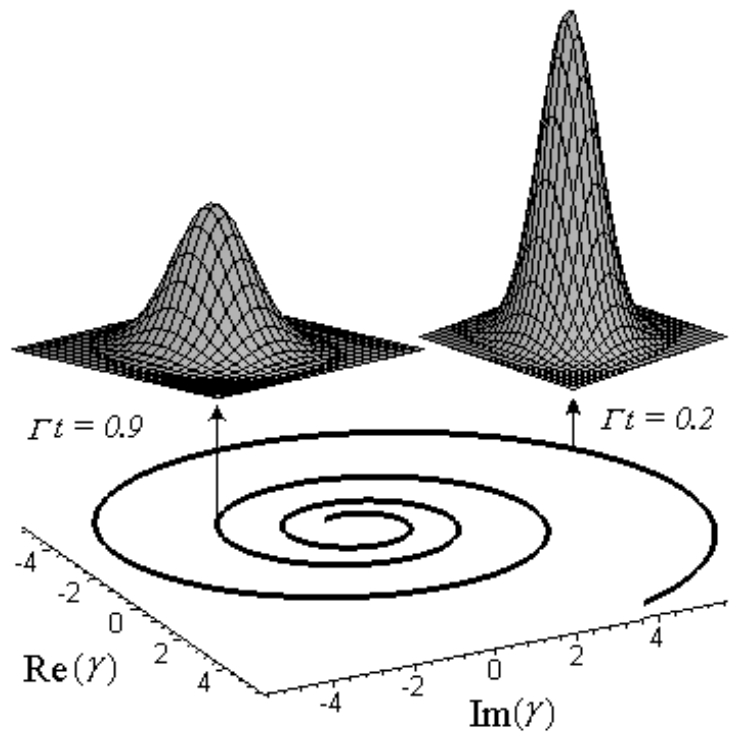
FIG. 2. Conditional distribution function $P(\gamma ,t)$ for an initially
prepared motional coherent state depicted at $\Gamma t=0.2$ and $0.9$. The
mean value of the amplitude moves on an exponential spiral displaying both
the effects due to coupling with the environment: the loss of excitation,
which carries the initial coherent state to the vacuum state, and the
diffusion due to nonzero temperature of the BG.

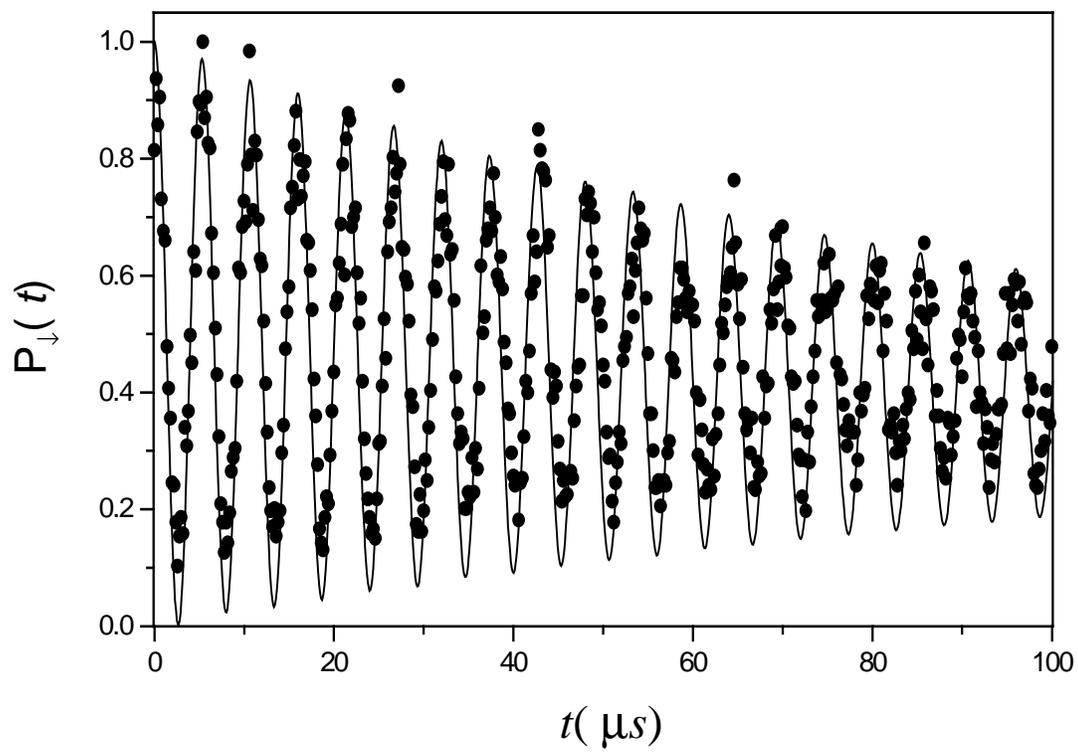
FIG. 3. Time evolution of the probability of fluorescence measurement of the
electronic ground state ${\sf P}_{\downarrow }(t)$, computed from our model
(full line) and measured in the NIST $^{9}$Be$^{+}$ experiments \cite{gatoW2}
(dots). The parameters used in the experiments $\eta _{L}=0.202$ and $\Omega
/2\pi \approx 475$ kHz, were adopted for the numerical calculation, together
with the values $\overline{n}_{th}=1.0$ and $\Gamma /g=$ $6.0\times $%
\thinspace $10^{-3}$ which are in good agreement with theoretical estimates.

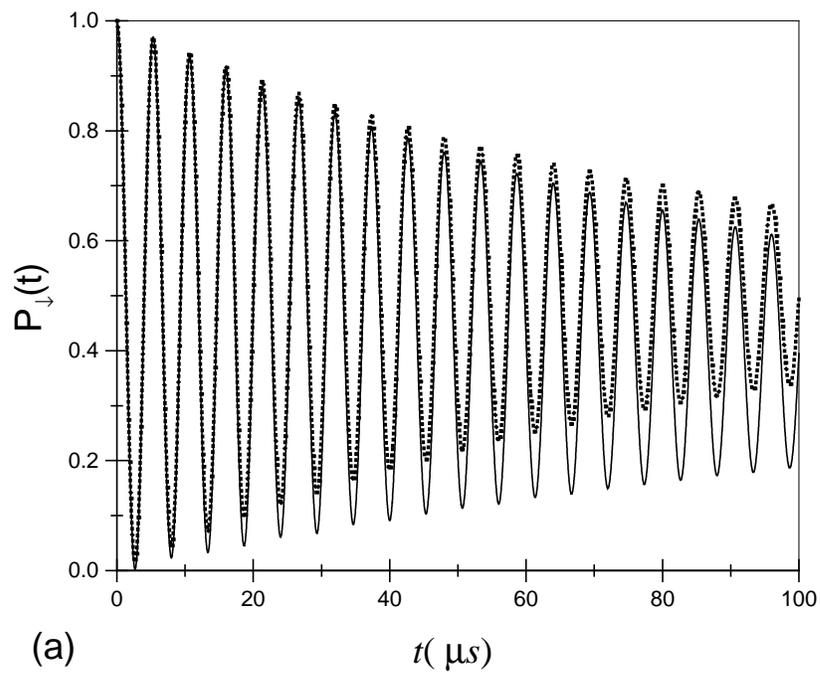
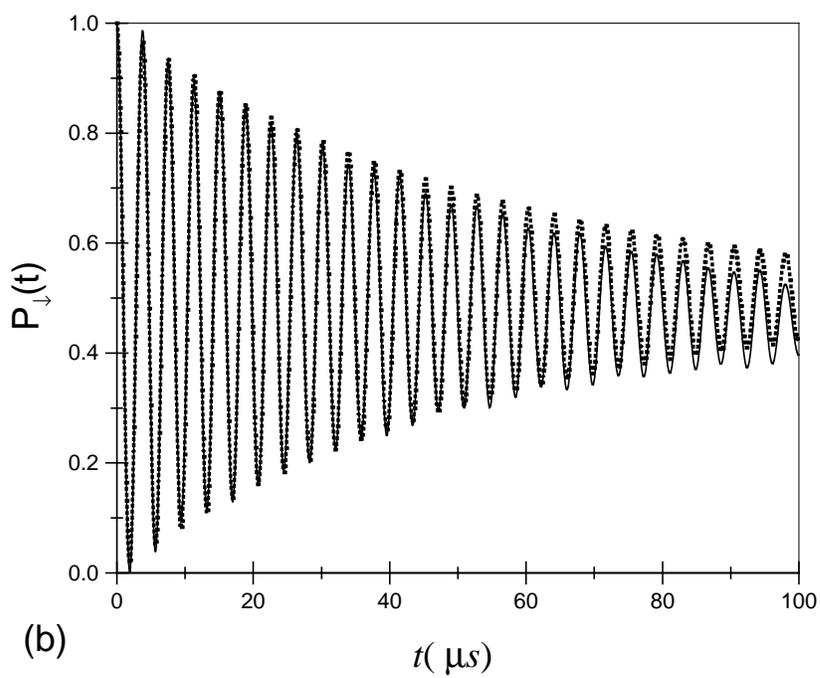
FIG. 4. Time evolution of the probability of fluorescence measurement of the
electronic ground state ${\sf P}_{\downarrow }(t)$ for the initially
prepared ionic states (a) $\left| n=0,\downarrow \right\rangle $ and (b) $%
\left| n=1,\downarrow \right\rangle $, according to our model (full lines)
and the heuristic relation (\ref{31}) (dotted lines), demonstrating the
measured sensitivity of the damping of Rabi oscillations to the motional
quantum number \cite{gatoW2}.

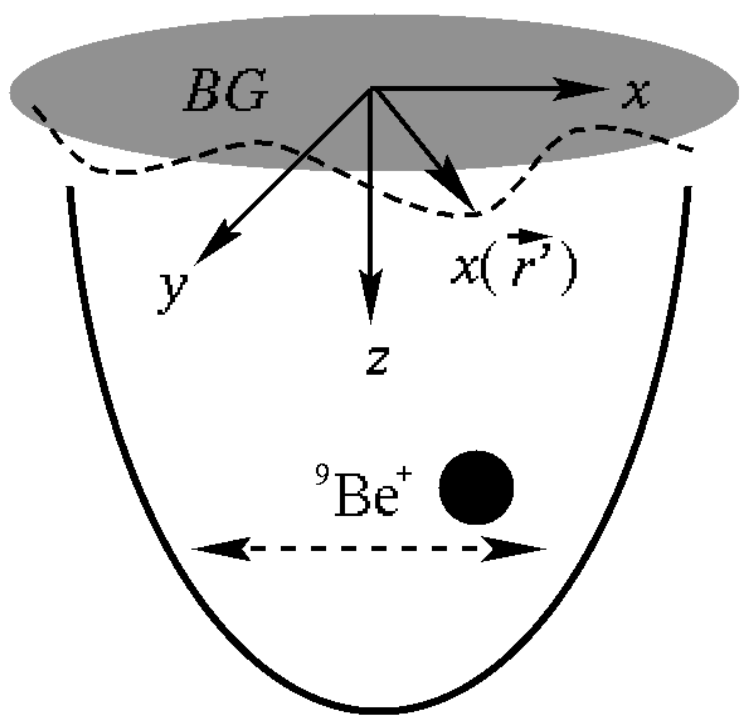
FIG. 5. Sketch of ion-BG interaction, showing the ion oscillating in the $x$%
-direction, assumed to be at a distance $z$ from the origin of the
coordinate system which is fixed in the center of the BG surface located in
the $xy$-plane. The ion is scattered by the oscillations of the BG surface,
whose dynamic roughness is given by $\xi \left( \overrightarrow{r^{\prime }}%
\right) $.

\end{document}